\documentclass[conference,10pt]{IEEEtran}
\IEEEoverridecommandlockouts

\usepackage{cite}
\usepackage{amsmath,amssymb,amsfonts}
\usepackage{algorithmic}
\usepackage{graphicx}
\usepackage{textcomp}
\usepackage{xcolor}
\usepackage{booktabs}
\usepackage{url}
\usepackage[hidelinks]{hyperref}
\usepackage{enumitem}
\usepackage{bbm}
\usepackage{microtype}
\usepackage{balance}

\begin{document}

\title{An End-to-End Hybrid Quantum--Classical Sampling Workflow for Discrete Markov Random Fields: A Reproducible Case Study}

\author{
\IEEEauthorblockN{Arul Rhik Mazumder}
\IEEEauthorblockA{\textit{School of Computer Science} \\
\textit{Carnegie Mellon University}\\
Pittsburgh, PA, USA \\
arulm@andrew.cmu.edu}
\and
\IEEEauthorblockN{Elias Towe}
\IEEEauthorblockA{\textit{Department of Electrical and Computer Engineering} \\
\textit{Carnegie Mellon University}\\
Pittsburgh, PA, USA \\
towe@andrew.cmu.edu}
}

\maketitle

\begin{abstract}
Amplitude encoding prepares a quantum state whose measurement outcomes are independent samples from a target distribution, so it has no autocorrelation cost. For a discrete Markov random field (MRF), however, preparing that state requires classically enumerating all $2^n$ probabilities, so no exponential quantum speedup is possible. We measure the practical value of sample independence in this regime. On 60 synthetic MRF instances ($n=8,10,12$ over five graph families) we compare an amplitude-encoded sampler with ideal shots against classical Markov chain Monte Carlo (MCMC) under a fixed protocol. Its effective-sample-size (ESS) advantage falls from $16.4\times$ against single-site Gibbs to $7.3\times$ against block Gibbs and $1.8\times$ against parallel tempering. Most of the apparent advantage therefore depends on the choice of baseline. Once the $2^n$ probabilities are available, exact inverse-CDF sampling from the same table also yields independent samples, and it does so faster than simulated quantum shots on the same CPU. A shallow hardware-efficient variational circuit falls well short of a matched-budget matrix product state (fidelity $0.17$ vs.\ $0.88$ at $n=12$), and clique-aware entanglement gives no measurable benefit over a linear layout. We do not claim a quantum advantage. Code and data are available at \url{https://github.com/arulrhikm/QuantumDGM}.
\end{abstract}


\section{Introduction}

\subsection{Motivation}

Markov random fields (MRFs) encode conditional independence through undirected graphs and are widely used in computer vision~\cite{li2009markov}, computational biology~\cite{kamisetty2013coevolution}, statistical physics and generative modeling~\cite{salakhutdinov2009deep}. The central computational task is drawing samples from the joint distribution
\begin{equation}
P_\theta(\mathbf{x}) = \frac{1}{Z(\theta)}\exp\!\left(\sum_{C\in\mathcal{C}}\sum_{\boldsymbol{y}\in\mathcal{X}_C}\theta_{C,\boldsymbol{y}}\,\phi_{C,\boldsymbol{y}}(\mathbf{x})\right),
\end{equation}
where $\mathcal{C}$ is the set of maximal cliques, $\phi_{C,\boldsymbol{y}}$ are indicator sufficient statistics and the partition function $Z(\theta)$ sums over all $2^n$ configurations of $n$ binary variables. Exact inference is therefore intractable for all but the smallest models, and many approximate methods exist.

Classical MCMC~\cite{geman1984stochastic} mixes slowly and produces correlated samples on loopy graphs. Variational inference~\cite{wainwright2008graphical} and loopy belief propagation~\cite{pearl1988probabilistic} have their own accuracy and convergence limits. The amplitude-encoding approach studied here does not avoid the $O(2^n)$ classical preprocessing. It generates i.i.d.\ samples only after $P_\theta$ has been enumerated. We ask how large the autocorrelation cost of classical MCMC is under a fixed protocol (\S\ref{sec:ess}) and whether any quantum property beyond i.i.d.\ output gives a practical advantage (\S\ref{sec:limitations_mcmc}).

\subsection{Benchmark Instances and Pipeline}

Our benchmark consists of synthetic MRFs on chain, barbell, barbell-path, Erd\H{o}s--R\'enyi and two-clique graphs at small to moderate $n$. Clique parameters are drawn as $\theta_{C,\boldsymbol{y}} \sim \mathcal{U}(-5,0)$, which gives peaked distributions (\S\ref{sec:marginals}) without entering the low-temperature regime where sampling becomes combinatorial optimization. Synthetic instances let us compute the probabilities of all $2^n$ configurations by direct enumeration, and hence exact fidelities, KL divergences and total variation distances.

We prepare states in one of two ways, depending on problem size:
\begin{enumerate}[leftmargin=*]
    \item \textbf{Amplitude encoding} for $n=8,10,12$, where the $2^n$ diagonal can be enumerated. Qiskit's \texttt{StatePreparation} primitive prepares a state whose squared amplitudes equal the target probabilities.
    \item \textbf{Variational quantum circuits (VQC)} for larger $n$. A hardware-efficient ansatz with linear, clique-aware or all-to-all entanglement compresses the target into a number of trainable parameters polynomial in $n$.
\end{enumerate}

\subsection{Evaluation Metrics}

We measure sampling quality with three metrics:
\begin{itemize}[leftmargin=*, topsep=0pt, partopsep=0pt, itemsep=2pt, parsep=0pt]
    \item \textbf{Fidelity} between an approximate or empirical distribution $\hat{P}$ and the target $P_\theta$,
    \[
    F(\hat{P}, P_\theta) = \Bigg(\sum_x \sqrt{\hat{P}(x)\,P_\theta(x)}\Bigg)^2 \in [0,1],
    \]
    which is the classical (Bhattacharyya) fidelity. It equals $1$ for identical distributions and $0$ for disjoint supports.
    \item \textbf{Kullback--Leibler (KL) divergence}
    \[
    D_{\mathrm{KL}}(\hat{P}\,\|\,P_\theta) = \sum_x \hat{P}(x)\log\frac{\hat{P}(x)}{P_\theta(x)},
    \]
    which is also the VQC training objective.
    \item \textbf{Total variation (TV) distance}
    \[
    \mathrm{TV}(\hat{P}, P_\theta) = \tfrac{1}{2}\sum_x \big|\hat{P}(x) - P_\theta(x)\big|,
    \]
    which equals the largest difference in probability that the two distributions assign to any event.
\end{itemize}
Fidelity is the primary metric in what follows. It is most sensitive to support alignment, while KL is sensitive to tail mismatch and TV to pointwise error.

\subsection{Scope and Contributions}

Our pipeline computes the target diagonal classically and then prepares $|\psi\rangle = \sum_x \sqrt{P_\theta(x)}|x\rangle$ with \texttt{StatePreparation} (\S\ref{sec:amp}). Our contributions are:
\begin{enumerate}[leftmargin=*]
    \item A comparison of ESS between amplitude-encoded i.i.d.\ sampling and four MCMC samplers (single-site Gibbs, block Gibbs, tuned-block Gibbs and parallel tempering) on 60 instances stratified by graph family (\S\ref{sec:ess}).
    \item A wall-clock comparison against exact inverse-CDF sampling, the natural classical baseline once the full distribution has been enumerated (\S\ref{sec:ess}).
    \item A three-seed MPS scaling study through $n=40$, scored against the exact $P_\theta$ at $n=12$ and against a $\chi=32$ MPS reference beyond (\S\ref{sec:scaling}).
    \item Open-source code and the result files behind every table and figure.
\end{enumerate}
We also report negative and null results. Shallow hardware-efficient VQC ans\"atze underperform matched-budget MPS at every measured $n\le12$ (\S\ref{sec:vqc-neg}). Clique-aware entanglement gives no measurable benefit over linear entanglement (\S\ref{sec:entanglement}). Mean-field approximation beats the VQC on full distributional fidelity at $n=8$ and $n=10$ (\S\ref{sec:classical_baselines}).

\section{Related Work}

Amin et al.~\cite{amin2018qbm} introduced quantum Boltzmann machines. Low, Yoder and Chuang~\cite{low2014bayes} studied quantum inference on Bayesian networks, and Verdon et al.~\cite{verdon2019vqe} prepared thermal states variationally. Layden et al.~\cite{layden2023qemcmc} and Ferguson and Wallden~\cite{ferguson2025qemcmc} use quantum proposals to accelerate MCMC.

Huggins et al.~\cite{huggins2019towards} relate parameterized circuits to tensor networks. Liu and Wang~\cite{liu2018differentiable} and Benedetti et al.~\cite{benedetti2019generative} train circuits as generative models whose output distribution is given by the Born rule. Wittek and Gogolin~\cite{wittek2017quantum} study quantum inference in Markov logic networks.

Piatkowski and Zoufal~\cite{piatkowski2024qcgms} construct quantum circuits for graphical models (QCGMs) that embed the model unitarily and sample by ancilla-based real-part extraction with repeat-until-success. Our encoder instead precomputes $P_\theta$ classically and prepares the state directly. This forgoes the potential speedup of Hamiltonian simulation but allows every stage to be checked against direct enumeration. Unlike the works above, we benchmark ESS, wall-clock time and fidelity for known discrete MRF distributions in the regime where $P_\theta$ can be precomputed.

\section{Methods}

\subsection{Amplitude Encoding}
\label{sec:amp}

For $n=8,10,12$ we construct the diagonal Hamiltonian
\begin{equation}
H_\theta = -\sum_{C\in\mathcal{C}}\sum_{\boldsymbol{y}} \theta_{C,\boldsymbol{y}}\,\Phi_{C,\boldsymbol{y}},
\end{equation}
where $\Phi_{C,\boldsymbol{y}}$ is a diagonal operator whose $(j,j)$ entry indicates whether configuration $\mathbf{x}^j$ agrees with assignment $\boldsymbol{y}$ on clique $C$. The target distribution is then
\[
P_\theta(\mathbf{x}^j) = \frac{(\exp(-H_\theta))_{j,j}}{\operatorname{Tr}(\exp(-H_\theta))}.
\]
The pipeline proceeds as follows:
\begin{enumerate}[leftmargin=*]
    \item Compute the diagonal of $H_\theta$ in $O(2^n)$ time by summing clique contributions.
    \item Exponentiate the diagonal entries to obtain unnormalized probabilities $\tilde{P}(x) = e^{-H_\theta(x,x)}$.
    \item Normalize to obtain the target probabilities $P_\theta(x) = \tilde{P}(x)/Z$.
    \item Compute the amplitudes $\alpha_x = \sqrt{P_\theta(x)}$.
    \item Reorder the amplitudes from the model's big-endian convention to Qiskit's little-endian bit ordering.
    \item Prepare $|\psi\rangle = \sum_x \alpha_x |x\rangle$.
    \item Measure in the computational basis to obtain samples from $\hat{P}$.
\end{enumerate}
Storing only the diagonal of $H_\theta$ takes $O(2^n)$ memory, whereas the dense Hamiltonian would take $O(4^n)$.

Because $P_\theta$ is precomputed, the pipeline can be audited end to end against direct enumeration. Any deviation from $F=1$ must then come from bit ordering, normalization or backend numerics. The $O(2^n)$ preprocessing is the same cost that any classical method working from the enumerated distribution incurs, including exact inverse-CDF sampling.

\subsection{Variational Compression}
\label{sec:vqc}

For larger $n$ the full diagonal of $H_\theta$ becomes impractical to store, and we instead train a hardware-efficient ansatz~\cite{kandala2017hardware}
\begin{equation}
U(\boldsymbol{\varphi}) = \prod_{l=1}^{d}\!\left[\!\left(\bigotimes_{i=1}^{n} R_Y(\varphi_{i,l})\right)\!\left(\prod_{(i,j)\in E}\!\mathrm{CX}_{i,j}\right)\!\right]\!,
\end{equation}
which consists of $d$ layers of single-qubit $R_Y$ rotations interleaved with CNOT gates on a fixed edge set $E$. This gives $nd$ trainable parameters (one angle per qubit per layer), far fewer than the $2^n$ amplitudes of the target distribution. Starting from $|0\rangle^{\otimes n}$, the circuit prepares $|\psi(\boldsymbol{\varphi})\rangle = U(\boldsymbol{\varphi})|0\rangle^{\otimes n}$ and induces the distribution $\hat{P}_{\boldsymbol{\varphi}}(x) = |\langle x|\psi(\boldsymbol{\varphi})\rangle|^2$.

We compare three choices of the edge set $E$:
\begin{enumerate}[leftmargin=*]
    \item \textbf{Linear:} CNOTs between adjacent qubits $(i,i+1)$. This uses $n-1$ gates per layer but has limited reach for long-range correlations.
    \item \textbf{Clique:} CNOTs that follow the cliques of the MRF graph and couple qubits that share a factor of the target distribution.
    \item \textbf{Full:} all-to-all CNOTs. This is the most expressive choice and takes roughly $3\times$ longer to train because the number of entangling gates per layer is quadratic in $n$.
\end{enumerate}

We train by minimizing
\[
\min_{\boldsymbol{\varphi}} \; D_{\mathrm{KL}}(\hat{P}_{\boldsymbol{\varphi}} \,\|\, P_\theta)
\]
with parameter-shift gradient estimates~\cite{schuld2019evaluating} and first-order updates. Training uses learning rate $0.05$ at depth $d=3$ for 30--80 iterations from fixed initialization seeds.

The clique factorization $P_\theta(\mathbf{x}) \propto \prod_C f_C(\mathbf{x}_C)$ confines MRF distributions to a low-dimensional subset of the $2^n$-dimensional probability simplex. Parameterized circuits can be viewed as tensor-network decompositions~\cite{huggins2019towards} in which single-qubit rotations set local marginals and entangling gates carry correlations between neighboring variables. An entanglement pattern that matches the graph (the clique strategy) might therefore be expected to compress MRFs well. Section~\ref{sec:entanglement} tests this expectation and finds no measurable benefit at the sizes studied.

\subsection{Variants That Did Not Help}

Several exploratory variants did not improve results:
\begin{enumerate}[leftmargin=*]
    \item \textbf{Deeper circuits ($d > 5$).} Additional layers gave small fidelity gains and longer training, consistent with harder optimization and possible barren plateaus~\cite{mcclean2018barren}. Initialization and optimizer choice matter more at larger depth.
    \item \textbf{Random entanglement.} Randomly chosen two-qubit edges gave less stable results across seeds than structured patterns.
    \item \textbf{Short training.} Small iteration budgets converged prematurely to materially lower fidelity. We therefore set the budget per problem size and graph family.
    \item \textbf{Squared-error loss.} An $\ell_2$ loss matched high-probability configurations but left almost no mass on the tail (mode collapse). We use the KL divergence throughout.
\end{enumerate}

\subsection{Experimental Setup}
\label{sec:setup}

We run Qiskit circuits on local Aer and on BlueQubit cloud simulators through a common adapter layer. All experiments use fixed random seeds, and the public repository contains the result files behind every reported number.

The ESS study compares the amplitude-encoded sampler with four MCMC samplers on the same 60 instances (five graph families at $n=8,10,12$). Random-scan single-site Gibbs updates one variable per step. Block Gibbs resamples a randomly chosen maximal clique at each step. Tuned-block Gibbs cycles deterministically through the two largest maximal cliques. On four of the five families these two cliques do not cover every variable, so the tuned-block chain is not ergodic there and its ESS ratio is not a valid mixing comparison. All Gibbs variants use 1{,}000 burn-in steps and keep 3{,}000 samples.

Parallel tempering (PT) runs $K=8$ replicas on a geometric inverse-temperature ladder $\beta \in [0.1, 1.0]$ and proposes replica swaps every $L_{\mathrm{swap}}=10$ single-site Gibbs sweeps. It uses 1{,}000 burn-in sweeps per replica and keeps 3{,}000 samples from the $\beta=1$ replica. Its wall-clock time includes all replicas.

Exact inverse-CDF sampling reuses the precomputed $P_\theta$, builds the cumulative sums $c_j=\sum_{i\le j}P_\theta(\mathbf{x}^i)$ once per instance and draws 3{,}000 samples by binary search. For both i.i.d.\ samplers we report a sample-only rate and an amortized rate that adds the $O(2^n)$ preprocessing. The amortized amplitude-encoded rate includes computing $P_\theta$, whereas the amortized inverse-CDF rate includes only building $c_j$ from a given $P_\theta$. The sample-only rates are therefore the directly comparable pair. The amplitude-encoded sampler is timed with ideal shots drawn from the exact output distribution of the prepared state, and all rates are measured on one local CPU.

ESS is computed on the Hamming-weight series of the retained samples using Geyer's initial positive sequence estimator~\cite{geyer1992practical}. For $N$ retained samples with integrated autocorrelation time $\tau$, the ESS is $N/\tau$.

\section{Results}

\subsection{Effective Sample Size and Sampling Rate}
\label{sec:ess}

\begin{figure}[t]
\centering
\includegraphics[width=\linewidth]{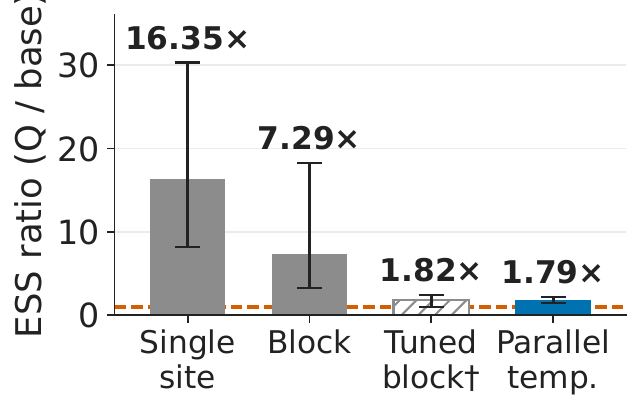}
\caption{ESS ratio of the amplitude-encoded sampler to each MCMC baseline. Bars are means over 60 instances and whiskers span the minimum and maximum. The dashed line marks a ratio of $1$. \dag\,Tuned-block Gibbs is not ergodic on four of the five families (\S\ref{sec:setup}).}
\label{fig:expD_baseline}
\end{figure}

Each shot of the amplitude-encoded sampler is an independent draw, so its integrated autocorrelation time is $\tau \approx 1$ on every graph. Table~\ref{tab:ess_baselines} and Fig.~\ref{fig:expD_baseline} compare its ESS with the four MCMC samplers over 60 matched instances (\S\ref{sec:setup}).

The ratios fall monotonically from $16.35\times$ against single-site Gibbs to $7.29\times$ against block Gibbs, $1.82\times$ against tuned-block Gibbs and $1.79\times$ against parallel tempering. The tuned-block ratio comes from a chain that is not ergodic on four of the five families, which leaves parallel tempering as the strongest valid baseline. Against it the amplitude-encoded sampler keeps a small autocorrelation advantage. The fall from $16\times$ to under $2\times$ shows how much of the apparent advantage depends on the choice of baseline.

Sampling rate is a separate question (Table~\ref{tab:ess_per_sec}). The amplitude-encoded sampler and exact inverse-CDF sampling both draw independent samples from the same enumerated table, and inverse-CDF sampling is the faster of the two ($1.94\times10^{7}$ vs.\ $1.45\times10^{7}$ ESS/s, sample-only). Every MCMC sampler is at least three orders of magnitude slower. The amortized column adds the cost of computing $P_\theta$ to the amplitude-encoded sampler alone, which explains its much larger gap.

\begin{table}[t]
\centering
\caption{ESS ratio of the amplitude-encoded sampler to each MCMC baseline over 60 instances from five graph families. Each chain uses a 1\,000-step burn-in and keeps 3\,000 samples. Per-instance ESS values are single-chain estimates, so the ratios are best read at the level of the mean and range.}
\label{tab:ess_baselines}
\begin{tabular}{lcccc}
\toprule
\textbf{Baseline} & \textbf{Mean} & \textbf{Median} & \textbf{Min} & \textbf{Max} \\
\midrule
Single-site Gibbs                  & 16.35 & 15.99 & 8.24  & 30.31 \\
Block Gibbs                        &  7.29 &  6.78 & 3.31  & 18.30 \\
Tuned-block Gibbs\textsuperscript{\dag} &  1.82 &  1.91 & 1.03  &  2.43 \\
Parallel tempering                 &  1.79 &  1.80 & 1.43  &  2.21 \\
\bottomrule
\end{tabular}
\par\smallskip
{\footnotesize \textsuperscript{\dag}Not ergodic on four of the five graph families (\S\ref{sec:setup}).}
\end{table}

\begin{table}[t]
\centering
\caption{ESS per second (mean over 60 instances) on one local CPU. Sample-only rates exclude all $O(2^n)$ preprocessing. Amortized rates include it, which for the amplitude-encoded sampler means computing $P_\theta$ and for inverse-CDF sampling means only building the cumulative sums.}
\label{tab:ess_per_sec}
\setlength{\tabcolsep}{3pt}
\begin{tabular}{lcc}
\toprule
\textbf{Method} & \textbf{Sample-only ESS/s} & \textbf{Amortized ESS/s} \\
\midrule
Amplitude-enc.\ (ideal) & $14{,}478{,}397$ & $487{,}706$ \\
Exact inverse-CDF       & $19{,}425{,}750$ & $17{,}683{,}598$ \\
Single-site Gibbs       & $10{,}290$ & $10{,}290$ \\
Block Gibbs             & $267$ & $267$ \\
Tuned-block Gibbs\textsuperscript{\dag} & $360$ & $360$ \\
Parallel tempering      & $2{,}080$ & $2{,}080$ \\
\bottomrule
\end{tabular}
\par\smallskip
{\footnotesize \textsuperscript{\dag}Not ergodic on four of the five graph families (\S\ref{sec:setup}).}
\end{table}

\subsection{Amplitude-Encoding Verification}

Table~\ref{tab:amplitude} checks the amplitude-encoding pipeline on chain graphs at $n=8,10,12$. All three runs reach $F=1.000$ and $\mathrm{TV}=0.000$ on the BlueQubit statevector simulator. A correct call to \texttt{StatePreparation} on a statevector simulator should give this result, and the check confirms that the bit ordering, normalization and diagonal-only computation are implemented correctly.

\begin{table}[t]
\centering
\caption{Amplitude encoding on chain graphs with Qiskit \texttt{StatePreparation}. The logical depth is $O(2^n)$ before hardware decomposition.}
\label{tab:amplitude}
\begin{tabular}{lcccc}
\toprule
\textbf{$n$} & \textbf{States} & \textbf{Fidelity} & \textbf{TV} & \textbf{Time (s)} \\
\midrule
8  & 256  & 1.000 & 0.000 &  2.92 \\
10 & 1024 & 1.000 & 0.000 &  3.91 \\
12 & 4096 & 1.000 & 0.000 & 12.71 \\
\bottomrule
\end{tabular}
\par\smallskip
{\footnotesize Backend: BlueQubit statevector simulator on all rows. Fidelities agree with $1$ to within $2\times10^{-7}$ (floating-point round-off).}
\end{table}

\subsection{Matched-Budget VQC and MPS}
\label{sec:vqc-neg}

\begin{figure}[t]
\centering
\includegraphics[width=\linewidth]{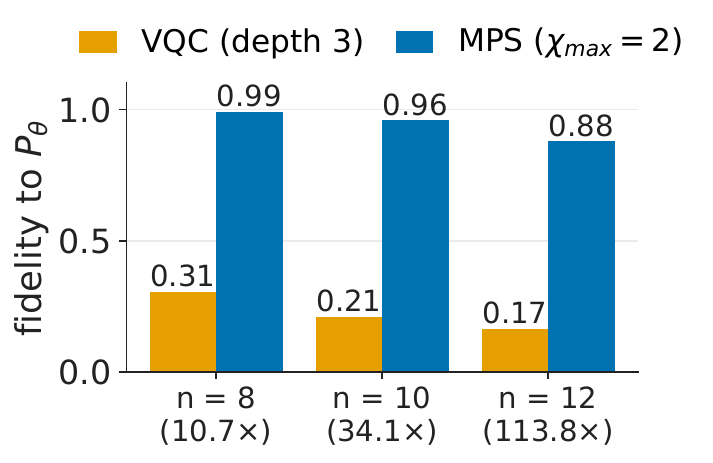}
\caption{Fidelity to $P_\theta$ of the matched-budget VQC and MPS from Table~\ref{tab:compression}. The label under each pair gives the VQC compression ratio.}
\label{fig:expB_headline}
\end{figure}

Table~\ref{tab:compression} and Fig.~\ref{fig:expB_headline} compare a depth-3 hardware-efficient VQC with an MPS of bond dimension $\chi_{\max}=2$ at $n=8,10,12$. The VQC fidelity is far below the MPS fidelity at every size. It falls from $0.306$ to $0.165$ while the MPS fidelity falls from $0.990$ to $0.878$. The VQC parameter counts give nominal compressions of $10.7\times$, $34.1\times$ and $113.8\times$, but no VQC fidelity exceeds $0.31$. The gap $F_{\mathrm{VQC}}-F_{\mathrm{MPS}}$ is non-monotone in $n$ and shows no sign of a crossover at accessible sizes. VQC training time grows with $n$ because of parameter-shift gradient evaluations, and MPS runs $2.5$--$3.2\times$ faster at matched sizes.

An SGD ablation at $n=12$ gives $F_{\mathrm{VQC}}=0.183$ against $F_{\mathrm{MPS}}=0.875$ at the same $\chi_{\max}=2$. Over 40 steps the KL loss falls by only $4.8\%$, which is consistent with limited depth-3 expressivity as the cause of the gap. Mean absolute partial derivatives stay between $0.03$ and $0.17$ at all measured sizes, which argues against a barren-plateau explanation~\cite{mcclean2018barren}.

These VQC fidelities are lower than those in Table~\ref{tab:classical_baselines} because the two experiments use different configurations. Table~\ref{tab:compression} fixes the ansatz at $nd$ parameters (\S\ref{sec:vqc}) for a matched comparison with $\chi_{\max}=2$ MPS, whereas Table~\ref{tab:classical_baselines} uses a depth-3 VQC without that cap.

\begin{table}[t]
\centering
\caption{Matched-budget VQC and MPS ($\chi_{\max}=2$). Targets and the VQC are evaluated on the BlueQubit statevector simulator and MPS on the BlueQubit MPS simulator. Compression is $2^n/n_{\mathrm{VQC\,params}}$, Gap is $F_{\mathrm{VQC}}-F_{\mathrm{MPS}}$ and $t$ is wall-clock time.}
\label{tab:compression}
\begin{tabular}{lcccccc}
\toprule
\textbf{$n$} & \textbf{$F_{\mathrm{VQC}}$} & \textbf{$F_{\mathrm{MPS}}$} & \textbf{Gap} & \textbf{Comp.} & \textbf{$t_{\mathrm{VQC}}$ (s)} & \textbf{$t_{\mathrm{MPS}}$ (s)} \\
\midrule
8  & 0.306 & 0.990 & $-0.684$ &  $10.7\times$ &  212 &  84 \\
10 & 0.210 & 0.958 & $-0.748$ &  $34.1\times$ &  320 & 101 \\
12 & 0.165 & 0.878 & $-0.713$ & $113.8\times$ &  421 & 133 \\
\bottomrule
\end{tabular}
\end{table}

\subsection{Entanglement Strategy}
\label{sec:entanglement}

Table~\ref{tab:delta_density} reports the entanglement sweep (108 runs). Clique entanglement gives no measurable advantage over linear entanglement ($\Delta F=-2.28\times10^{-4}$), and the clique-minus-linear difference does not correlate with graph density ($R^2=6.6\times10^{-4}$). Full entanglement gives a small gain ($\Delta F=+2.58\times10^{-3}$) at $5\times$ the CX gate count.

\begin{table}[t]
\centering
\setlength{\tabcolsep}{4pt}
\caption{Entanglement strategies over 108 runs (3 strategies, 4 families, 3 seeds and $n\in\{10,12,14\}$). Mean fidelity is over matched (family, $n$, seed) triples. $\Delta F$ is the mean difference from linear, and $R^2$ is from an OLS regression of $\Delta F_{\text{clique}-\text{linear}}$ on graph density.}
\label{tab:delta_density}
\begin{tabular}{lccc}
\toprule
\textbf{Statistic} & \textbf{Linear} & \textbf{Clique} & \textbf{Full} \\
\midrule
Mean fidelity (36 runs each)   & 0.217 & 0.217 & 0.220 \\
$\Delta F$ vs.\ linear         & ---   & $-2.28\times10^{-4}$ & $+2.58\times10^{-3}$ \\
$R^2$ ($\Delta F$ vs.\ density) & ---  & $6.6\times10^{-4}$ & --- \\
\bottomrule
\end{tabular}
\par\smallskip
{\footnotesize Families: barbell, chain, Erd\H{o}s--R\'enyi and two-clique.}
\end{table}

\subsection{Comparison with Classical Approximations}
\label{sec:classical_baselines}

Table~\ref{tab:classical_baselines} compares the unconstrained depth-3 VQC with mean-field approximation and loopy belief propagation at $n=6,8,10$. Mean-field matches the global distribution better than the VQC at $n=8$ ($0.787$ vs.\ $0.664$) and $n=10$ ($0.606$ vs.\ $0.556$). The VQC has the lowest singleton-marginal error of the three methods at every size, so it reproduces individual marginals more accurately than its global fidelity suggests.

\begin{table}[t]
\centering
\caption{VQC and classical approximations against the exact $P_\theta$. Fidelity rows compare full distributions, and the singleton rows give the mean absolute error of per-variable marginals. Amplitude encoding is included for reference.}
\label{tab:classical_baselines}
\begin{tabular}{lccc}
\toprule
\textbf{Method} & \textbf{$n=6$} & \textbf{$n=8$} & \textbf{$n=10$} \\
\midrule
\multicolumn{4}{l}{\textit{Full distributional fidelity}} \\
VQC (depth 3)              & 0.744 & 0.664 & 0.556 \\
Mean-field                 & 0.710 & 0.787 & 0.606 \\
Amplitude encoding         & 1.000 & 1.000 & 1.000 \\
\midrule
\multicolumn{4}{l}{\textit{Singleton marginal MAE (lower is better)}} \\
VQC (depth 3)              & 0.109 & 0.096 & 0.111 \\
Mean-field                 & 0.149 & 0.124 & 0.156 \\
Loopy BP                   & 0.145 & 0.124 & 0.156 \\
\bottomrule
\end{tabular}
\end{table}

\subsection{Distribution Hardness}
\label{sec:marginals}

Table~\ref{tab:fidelity_decomp} shows that the benchmark distributions are highly peaked. Their entropy is far below $n$ bits, and at $n=12$ the top modes hold more than $99.9\%$ of the mass. A shallow VQC must place almost all of its mass on these few configurations, which makes such targets demanding.

\begin{table}[t]
\centering
\caption{Distribution hardness (3 seeds per $n$, arithmetic means). Head mass is the fraction of probability mass in the top modes, entropy is in bits and max prob is the probability of the single most likely configuration.}
\label{tab:fidelity_decomp}
\begin{tabular}{lccc}
\toprule
\textbf{$n$} & \textbf{Mean head mass} & \textbf{Mean entropy (bits)} & \textbf{Mean max prob} \\
\midrule
 8 & 0.977 & 2.93 & 0.351 \\
10 & 0.997 & 3.55 & 0.277 \\
12 & 1.000 & 4.26 & 0.231 \\
\bottomrule
\end{tabular}
\par\smallskip
{\footnotesize At $n=12$ the distributions are concentrated on $\lesssim16$ of the $4096$ configurations.}
\end{table}

\subsection{MPS Scaling}
\label{sec:scaling}

\begin{figure}[t]
\centering
\includegraphics[width=\linewidth]{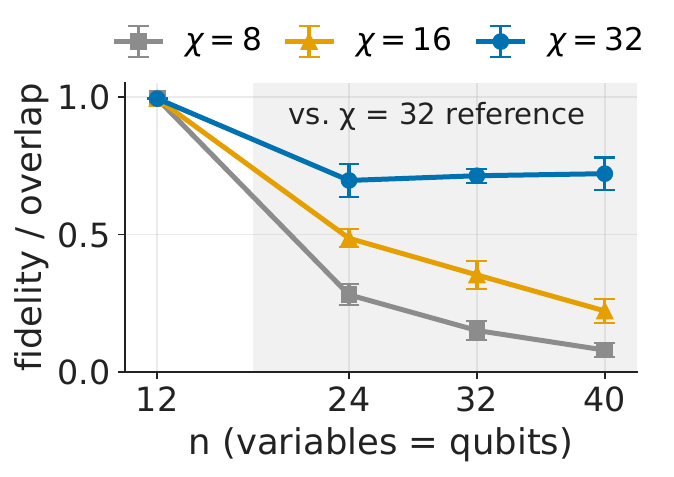}
\caption{MPS fidelity at $n=12$ and overlap with a $\chi=32$ reference for $n\ge24$ (shaded), with mean and 95\% CI over 3 seeds. The $\chi=32$ points in the shaded region are self-overlaps set by shot noise.}
\label{fig:expB_mpsvschi}
\end{figure}

We run MPS with three seeds for each $n\in\{12,24,32,40\}$ and $\chi\in\{8,16,32\}$ on the BlueQubit MPS simulator (Table~\ref{tab:mps_scaling}, Fig.~\ref{fig:expB_mpsvschi}). At $n=12$ the MPS output is scored against the exact $P_\theta$, and every bond dimension reaches $F\ge0.993$. For $n\ge24$ enumeration is infeasible, so we score each run by its finite-shot overlap with samples from a $\chi=32$ MPS. These values measure agreement with $\chi=32$ and do not measure fidelity to $P_\theta$. The $\chi=32$ entries ($0.70$--$0.72$) are self-overlaps limited by shot noise, and the $\chi=8$ and $\chi=16$ entries fall steadily with $n$ relative to that level.

\begin{table}[t]
\centering
\caption{MPS scaling (3 seeds per point, mean and 95\% CI). Compression is $2^n/n_{\mathrm{MPS\,params}}$. At $n=12$ the entry is the fidelity to the exact $P_\theta$. For $n\ge24$ it is the finite-shot overlap with a $\chi=32$ MPS reference.}
\label{tab:mps_scaling}
\begin{tabular}{lcc p{2.3cm} c}
\toprule
\textbf{$n$} & \textbf{$\chi$} & \textbf{$F$ or overlap} & \textbf{Compression} & \textbf{Trials} \\
\midrule
12 &  8 & $0.9936 \pm 0.0019$ &           $3.12$ & 3 \\
12 & 16 & $0.9940 \pm 0.0009$ &           $0.79$ & 3 \\
12 & 32 & $0.9939 \pm 0.0015$ &           $0.20$ & 3 \\
\midrule
24 &  8 & $0.2805 \pm 0.0384$ &        $5{,}891$ & 3 \\
24 & 16 & $0.4860 \pm 0.0321$ &        $1{,}481$ & 3 \\
24 & 32 & $0.6963 \pm 0.0602$ &          $371$ & 3 \\
\midrule
32 &  8 & $0.1508 \pm 0.0359$ &  $1{,}109{,}237$ & 3 \\
32 & 16 & $0.3527 \pm 0.0519$ &    $278{,}460$ & 3 \\
32 & 32 & $0.7134 \pm 0.0256$ &     $69{,}760$ & 3 \\
\midrule
40 &  8 & $0.0793 \pm 0.0259$ & $224{,}573{,}453$ & 3 \\
40 & 16 & $0.2218 \pm 0.0433$ &  $56{,}327{,}440$ & 3 \\
40 & 32 & $0.7209 \pm 0.0588$ &  $14{,}104{,}983$ & 3 \\
\bottomrule
\end{tabular}
\par\smallskip
{\footnotesize Backend: BlueQubit MPS simulator on all rows.}
\end{table}

\section{Discussion}

Classical MCMC~\cite{brooks2011handbook} produces correlated samples through a Markov kernel, whereas each shot of the amplitude-encoded sampler is an independent draw from $|\psi\rangle$. This independence is the source of the sampler's ESS advantage. Section~\ref{sec:ess} shows that the advantage shrinks to under $2\times$ against parallel tempering and that inverse-CDF sampling from the same table is faster. The clique factorization of MRFs is a low-dimensional structure that tensor-network-like circuits could in principle exploit~\cite{huggins2019towards}. MPS exploits it at small $n$, while the shallow hardware-efficient circuits we tested do not (\S\ref{sec:vqc-neg} and \S\ref{sec:entanglement}).

\subsection{Limitations and Method Selection}
\label{sec:limitations_mcmc}

The amplitude-encoding pipeline needs $O(2^n)$ classical preprocessing, and we evaluate it only on simulators. We implement neither the full QCGM construction~\cite{piatkowski2024qcgms} nor noisy hardware. The sampling rates in Table~\ref{tab:ess_per_sec} are for ideal shots on one CPU. The fidelity results come from BlueQubit statevector and MPS simulators, which we did not cross-check on an independent backend. The tuned-block baseline (\S\ref{sec:setup}), the $n\ge24$ MPS reference (\S\ref{sec:scaling}) and the amortized rates (Table~\ref{tab:ess_per_sec}) carry the caveats stated where they are reported.

Within this scope we suggest inverse-CDF sampling when enumeration is feasible. Amplitude encoding is appropriate when independent quantum samples or a coherent state for later processing (as in amplitude estimation) is required. When $n$ is too large to enumerate, parallel tempering or block MCMC are the natural choices, and junction-tree methods suit low-treewidth graphs~\cite{lauritzen1988local}. We do not recommend depth-3 hardware-efficient VQCs for full-distribution sampling.

\subsection{Future Directions}

Natural extensions are a full QCGM implementation~\cite{piatkowski2024qcgms}, noisy-hardware runs with error mitigation, variational objectives that match marginals and hybrid quantum-classical sampling of distribution tails.

\section{Conclusion}

We compared an amplitude-encoded sampler, whose shots are independent, with four MCMC baselines on 60 discrete MRF instances. Its ESS advantage falls from $16.35\times$ against single-site Gibbs to $1.79\times$ against parallel tempering, so most of the apparent quantum advantage depends on the baseline. Inverse-CDF sampling from the same enumerated table is faster than the amplitude-encoded sampler, which leaves no wall-clock advantage in the regime where amplitude encoding applies. A shallow hardware-efficient VQC falls well short of matched-budget MPS at every $n\le12$ we tested. We do not claim a quantum advantage. Full QCGM Hamiltonian simulation, deeper or problem-aware ans\"atze and noisy-hardware benchmarks at scale are the most plausible routes to one. Code and data are available at \url{https://github.com/arulrhikm/QuantumDGM} under the MIT license.


\section*{AI Assistance}
Large language model assistance (Claude, Anthropic) was used for manuscript editing, structural revision and consistency checking. All experimental design, code, data and analysis are the authors' own work.

\section*{Acknowledgments}
This material is based on work supported in part by the National Science Foundation (NSF) under Award No. 2310949. It was conducted within the Center for Quantum Computing and Information Technologies (QCiT), an industry-university collaborative research center at Carnegie Mellon University. We also thank the open-source quantum computing community for the tools that enabled this work.

\balance
\bibliographystyle{IEEEtran}
\bibliography{references}

\end{document}